# Atom-photon, two-mode entanglement and two-mode squeezing in the presence of cross-Kerr nonlinearity

**Ali Mortezapour**[a*], **Mohammad Mahmoudi**[b] and **M R H Khajehpour**[c]

[a] Department of Physics, University of Guilan, P. O. Box 41335–19141, Rasht, Iran
[b] Physics Department, University of Zanjan, P.O. Box 45195-313, Zanjan, Iran
[c] Institute for Advanced Studies in Basic Sciences, P.O. Box 45195-159, Zanjan, Iran
Corresponding author. E-mail: mortezapour@guilan.ac.ir

**Abstract**:
The interaction of two quantized fields and three-level quantum system in a $\Lambda$-type configuration is investigated in the presence of cross-Kerr nonlinearity. We consider three models of coupling for the atom-photon interaction. First, we study the dynamical behavior of the atom-photon entanglement and show that increasing the cross-Kerr nonlinearity results in different behaviors in three considered models. Moreover, it is demonstrated that the two quantized modes can be entangled, on the other hand, by applying a classical driving field to the lower levels. Increasing the classical driving field destroys the long time atom-photon entanglement. Our results show that an oscillatory two-mode squeezing can be generated in the absence of a driving classical field and the cross-Kerr nonlinearity.

## 1. Introduction

Quantum mechanics is a fundamental and accurate theory that can show many amazing properties which are not describable by classical mechanics. Entanglement is one of the most striking of these properties which is contrary to intuitive expectation. Generally, a system consisting of two sub-systems is said to be entangled if its quantum state cannot be described as a simple product of the quantum states of the constituent sub-systems. Physically, it means that a measurement on one of the sub-systems yields information about the other sub-system. During the last few years, it has become clear that the importance of the phenomenon extends well beyond fundamental questions of quantum theory. Indeed, entanglement lies at the heart of many quantum-information processing tasks such as quantum teleportation [1], quantum cryptography [2], and dense coding [3].

Up to now, various schemes have been proposed to generate and detect a desirable entangled state in atomic-molecular-optical systems [4-9]. All of the entangling schemes rely on direct or indirect interactions between subsystems. Such interactions are almost ubiquitous in atomic systems. For instance, in a typical cavity QED system, coherently interaction of a two-level atom with a single quantized mode of an electromagnetic field in a cavity which is described by Jaynes-Cummings model (JCM) [10] directly leads to an atom-photon entanglement. It is worth noting that atom-photon entanglement has found application in new crucial quantum concepts such as quantum repeater [11, 12], quantum networks and quantum memories [13].

Due to these practical applications and noting the generalization of JCM in the cases of few-level atoms [14], intensity-dependent couplings [15,16] and nonlinearities [17, 18], the study of atom-photon entanglement in these systems have attracted attentions of many researchers. For example, atom-photon entanglement in the interaction between a V-type three-level atom and single-mode cavity field in a Kerr medium with intensity-dependent coupling has been studied by Obada et al. [19]. In addition, atom-photon entanglement in a $\Lambda$-type [20, 21] and cascade [22] three-level atom interacting with a non-correlated two-mode cavity field has been investigated in the presence of Kerr nonlinearity.

Furthermore, in the context of semi-classical theory of atom-light interaction, it has been shown that the induced steady state entanglement between an atom and its spontaneous emission can be controlled by intensity and detuning parameter of an applied field [23]. It has, also, been demonstrated that the atom-photon entanglement depends on the relative phase of applied fields [24, 25] and incoherent driving pump field [26].

In a more general system, when an atom interacts with a two-mode cavity field, we may also have a two-mode continuous variable entanglement. In other words, the quadrature operators of two-cavity modes can be entangled indirectly via atom-field interaction. Entanglement criteria for continuous variables have been worked out by Simon [27], Duan et al. [28], Hillery and Zubairy [29] and others. Based on these entanglement criteria, some experimental [7-9] and theoretical [30-35] works have been designed to create two-mode continuous variable entanglement. Some other researchers have studied squeezing properties of two-mode cavity radiation as well as its entanglement and have tried to find relations between these physical phenomena [36-38]. Over the last few years, several experiments have been conducted in the context of quantum information theory such as unconditional quantum teleportation [39-41], dense coding [42] and quantum key distribution [43] with the aid of two-mode continuous optical variables.

In this paper, similar to references [20, 21], we consider a $\Lambda$-type three-level atom that interacts with a non-correlated two-mode quantized field in a lossless cavity in the presence both Kerr and cross-Kerr nonlinearity. The atom-field coupling is assumed to possess some admissible intensity dependence. It is noteworthy that the novelty of this work relies on the fact that, in the previous similar works [20, 21], the authors did not consider cross-Kerr nonlinearity term [44, 45] in the hamiltonian and merely investigated atom-photon entanglement. But here, we incorporate cross-Kerr nonlinearity term in the hamiltonian and apply a classical field to the lower levels of the atom. Furthermore, we obtain the state vector of the system in a distinguished analytical method and study the dynamical behavior of the two-mode entanglement and the two-mode squeezing in addition to the atom-photon entanglement. In contrast to the case of the atom-photon entanglement, we examine its dynamical behavior, first for different initial atomic states and then in the case where only the cross-Kerr nonlinearity is present. Note that such results do not appear in [22, 23].

We show that the cross-Kerr effect destroys the two-mode entanglement and the two-mode squeezing in the system. Therefore, an interesting result of this paper is that it states that, in order to experimentally obtain the two-mode entanglement or the two-mode squeezing in such system, one has to avoid the cases in which the cross-Kerr nonlinearity might occur.

This paper is organized as follows: in section 2 we introduce the Hamiltonian of the system. It is, then, exactly solved and the state vectors are obtained. In section 3, the von Neumann entropy and the reduced atomic entropy are calculated. In addition, the effect of different parameters related to the atoms and to the fields as well as their influences on the dynamical behavior of the degree of atom-photon entanglement are studied. Furthermore, we study the two-mode entanglement and the two-mode squeezing in sections 4 and 5, respectively. Finally, some conclusions and comments are presented in section 6.

## 2. Model and Solutions

Consider a $\Lambda$-type three-level atom whose eigenstates are denoted as $|1\rangle, |2\rangle, |3\rangle$ correspond corresponding to the energies $\hbar\omega_1$, $\hbar\omega_2$ and $\hbar\omega_3$ ($\hbar\omega_3 < \hbar\omega_2 > \hbar\omega_1$). The atom interacts with a two-mode quantized cavity field which carries frequencies $\omega_L$ and $\omega_R$ in the presence of nonlinearities and a classical driving field of Rabi frequency $\Omega$, as shown in figure 1. The $|1\rangle \leftrightarrow |2\rangle$ transition is coupled to the left-mode of quantized field of the frequency $\omega_L$ and the right-mode of the frequency $\omega_R$ is coupled to the transition $|3\rangle \leftrightarrow |2\rangle$. The classical driving field of Rabi frequency $\Omega$ drives transition $|1\rangle \leftrightarrow |3\rangle$. For the sake of simplicity, we would assume that $\hbar = 1$.

The Hamiltonian in the rotating wave approximation is of the form:

$$H = H_A + H_F + H_{in}, \tag{1}$$

where $H_A$ is the free atomic part of the Hamiltonian,

$$H_A = \sum_{i=1}^{3} \omega_i |i\rangle\langle i|, \qquad (2)$$

and $H_F$ consists of a general free two-mode quantized part and a field-field contribution due to the nonlinear medium

$$H_F = \sum_{j=L,R} [\omega_j \hat{a}_j^+ \hat{a}_j + R_j(\hat{a}_j^+ \hat{a}_j)] + R_C(\hat{n}_R, \hat{n}_L). \qquad (3)$$

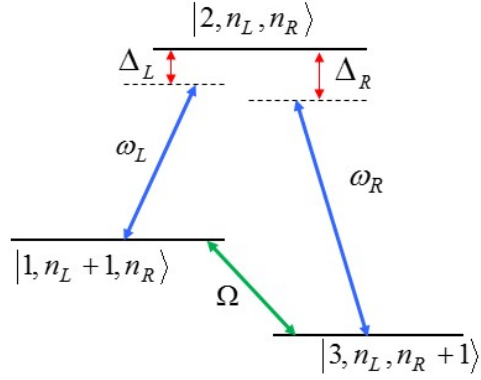

Figure1

The operators $\hat{a}_j^+$ and $\hat{a}_j$ ($j = R, L$) are the creation and annihilation operators for the $j$th mode of the cavity quantized field, respectively. These operators obey Boson commutation relation $[\hat{a}_j, \hat{a}_j^+] = 1$. $R_j(\hat{a}_j^+ \hat{a}_j)$ is the $j$th mode cavity quantized field nonlinearity due to Kerr effect usually written as $R_j(\hat{a}_j^+ \hat{a}_j) = \chi_j n_j(n_j - 1)$, where $\hat{n}_j = \hat{a}_j^+ \hat{a}_j$ is the number operator and $\chi_j$ is the self-action coefficient for the mode $j$ and $j = R, L$. The term $R_C(\hat{n}_R, \hat{n}_L) = \chi_c \hat{a}_R^+ \hat{a}_R \hat{a}_L^+ \hat{a}_L$ in the $H_F$ describes the cross-Kerr effect (photon-photon scattering between two incoming photons) which plays an important role in quantum non-demolition measurements [46, 47]. The coefficients $\chi_j$ are all proportional to the third-order susceptibility of the nonlinear medium. Finally, $H_{in}$ consists of all atom-fields interactions:

$$H_{in} = \lambda_L \hat{a}_L f(\hat{a}_L^+ \hat{a}_L)|2\rangle\langle 1| + \lambda_R \hat{a}_R f(\hat{a}_R^+ \hat{a}_R)|2\rangle\langle 3| + \Omega|1\rangle\langle 3| + h.c, \qquad (4)$$

where $\Omega$ is Rabi frequency of the classical driving field which couples two lower levels of the atom and induces coherence between them. Moreover, $\lambda_j$ is the vacuum Rabi frequency of the mode $j$ and $f_j(\hat{a}_j^+ \hat{a}_j)$ is a Hermitian function of $\hat{a}_j^+ \hat{a}_j$ which represents the intensity-dependent atom-field coupling. Here we assume,

$$f(\hat{a}_j^+ \hat{a}_j) = (\hat{a}_j^+ \hat{a}_j)^\beta, \quad \beta = -1/2, 0, 1/2. \tag{5}$$

The spatial structure of the cavity field determines the value of $\beta$. Namely, arbitrary atom-field couplings ($\beta$) can be obtained by proper choice of the cavity mode structure. However, $\beta = 0$ means that the atom-field coupling is independent of the intensity of the field as in the case of JCM [10]. Note that $\beta = 1/2$ and $\beta = -1/2$ depict the intensity dependent models which show that atom-field coupling depends on the intensity of the field. We mention that the intensity dependent model with $\beta = 1/2$ ($\beta = -1/2$) was initially introduced by Buck and Sukumar in [15] (by Obada et al in [19]).

To derive the state vector of the system we apply the method used by Yoo and J. H. Eberly [14]. We define three constant operators of motion: the total electron number operator $\hat{P}_E$

$$\hat{P}_E = |1\rangle\langle 1| + |2\rangle\langle 2| + |3\rangle\langle 3|, \tag{6}$$

taken to be $\hat{P}_E = 1$ and the excitation number operators $\hat{N}_1$, $\hat{N}_2$ defined as

$$\hat{N}_1 = \hat{a}_L^+ \hat{a}_L - |1\rangle\langle 1| + \hat{I}, \quad \hat{N}_2 = \hat{a}_R^+ \hat{a}_R - |3\rangle\langle 3| + I. \tag{7}$$

We write the Hamiltonian H as the sum of two parts $H_I$ and $H_{II}$ where $H_I$ contains only $\hat{P}_E$, $\hat{N}_1$ and $\hat{N}_2$ operators, hence a constant of motion:

$$H = H_I + H_{II}, \quad \text{and} \quad [H_I, H_{II}] = 0, \tag{8}$$

$$H_I = \omega_L \hat{N}_1 + \omega_R \hat{N}_2 + \omega_2 \hat{P}_E - (\omega_L + \omega_R)\hat{I}, \tag{9}$$

$$H_{II} = -\Delta_L |1\rangle\langle 1| - \Delta_R |3\rangle\langle 3| + R_L(\hat{a}_L^+ \hat{a}_L) + R_R(\hat{a}_R^+ \hat{a}_R) + R_C(\hat{n}_L, \hat{n}_R) + H_{in}. \tag{10}$$

Here $\Delta_L = \omega_2 - \omega_1 - \omega_L$, $\Delta_R = \omega_2 - \omega_3 - \omega_R$ are detunings of two modes. Since $H_I$ and $H_{II}$ commute, the time evolution operator can be factorized as:

$$\hat{U}(t) = \exp(-iHt) = \hat{U}_I(t)\hat{U}_{II}(t), \tag{11}$$

$$\hat{U}_I(t) = \exp(-iH_I t), \quad \hat{U}_{II}(t) = \exp(-iH_{II}t). \tag{12}$$

For the states $|1, n_L + 1, n_R\rangle$, $|2, n_L, n_R\rangle$ and $|3, n_L, n_R + 1\rangle$, $\hat{N}_1$ and $\hat{N}_2$ are both constants. These states form a subspace which we call ($N_1, N_2$)-subspace. In the ($N_1, N_2$)-subspace the matrix representation of $H_I$ and $H_{II}$ are

$$H_I = (\omega_L \hat{N}_1 + \omega_R \hat{N}_2 + \omega_2 \hat{P}_E - (\omega_L + \omega_R)).\hat{I}, \tag{13}$$

$$H_{II} = \begin{bmatrix} r_1 & V_L^* & \Omega \\ V_L & r_2 & V_R \\ \Omega^* & V_R^* & r_3 \end{bmatrix} \tag{14}$$

Where:

$$r_1 = -\Delta_L + R_L(n_L+1) + R_R(n_R) + R_C(n_L+1, n_R),$$
$$r_2 = R_L(n_L) + R_R(n_R) + R_C(n_L, n_R),$$
$$r_3 = -\Delta_R + R_L(n_L) + R_R(n_R+1) + R_C(n_L, n_R+1), \qquad (15)$$
$$V_L = \lambda_L f(n_L+1)\sqrt{n_L+1},$$
$$V_R = \lambda_R f(n_R+1)\sqrt{n_R+1},$$

The eigenvalues of $\hat{H}_{II}$ are obtained by solving $\det|\hat{H}_{II} - E\hat{I}| = 0$.

$$E_i = -\frac{Y_1}{3} + \frac{2}{3}(\sqrt{Y_1^2 - 3Y_2})\mathrm{Cos}\left(\frac{1}{3}\mathrm{ArcCos}\left[\frac{9Y_1Y_2 - 2Y_1^3 - 27Y_3}{2(Y_1^2 - 3Y_2)^{\frac{3}{2}}}\right] + (i-1)\right), \quad i=1,2,3 \quad (16)$$

where

$$Y_1 = -(r_1 + r_2 + r_3),$$
$$Y_2 = r_1 r_2 + r_1 r_3 + r_2 r_3 - (|V_L|^2 + |V_R|^2 + |\Omega|^2), \qquad (17)$$
$$Y_3 = r_1|V_R|^2 + r_2|\Omega|^2 + r_3|V_L|^2 - (r_1 r_2 r_3 + V_L V_R^* \Omega + V_L^* V_R \Omega^*),$$

The eigenvectors are

$$|\psi_i\rangle = \begin{pmatrix} a_i \\ b_i \\ c_i \end{pmatrix}, \text{ with } i=1,2,3 \text{ and}$$

$$a_i = [V_L^* + V_R^*\Omega/(E_i - r_3)]/h_i,$$
$$b_i = [(E_i - r_1) - |\Omega|^2/(E_i - r_3)]/h_i,$$
$$c_i = [(V_L^*\Omega + V_R^*(E_i - r_1))/(E_i - r_3)]/h_i, \qquad (18)$$
$$h_i = [(V_L^* + V_R^*\Omega/(E_i - r_3))^2 + ((E_i - r_1) - |\Omega|^2/(E_i - r_3))^2$$
$$\qquad + ((V_L^*\Omega + V_R^*(E_i - r_1))/(E_i - r_3))^2]^{1/2}$$

So the operator $H_{II}$ is diagonalized by the matrix,

$$V = \begin{bmatrix} a_1 & a_2 & a_3 \\ b_1 & b_2 & b_3 \\ c_1 & c_2 & c_3 \end{bmatrix}, \qquad (19)$$

as

$$V^+ H_{II} V = \begin{bmatrix} E_1 & 0 & 0 \\ 0 & E_2 & 0 \\ 0 & 0 & E_3 \end{bmatrix}. \qquad (20)$$

Unitarity of $\hat{V}$ leads to the following matrix representation of $\hat{U}_{II}(t)$ in the bare state space:

$$\hat{U}_{II}(t) = \exp(-iH_{II}t) = V\exp(-i(V^+H_{II}V)t)V^+ = V\begin{bmatrix} \exp(-iE_1t) & 0 & 0 \\ 0 & \exp(-iE_2t) & 0 \\ 0 & 0 & \exp(-iE_3t) \end{bmatrix}V^+.$$

(21)

On the other hand, $\hat{U}_I(t)$ is just a phase factor:

$$\hat{U}_I(t) = e^{-i(\omega_L N_1 + \omega_R N_2 + \omega_2 - (\omega_L + \omega_R))t} \times \hat{I}, \tag{22}$$

The initial state $|\Psi(0)_{AF}\rangle$ of the combined atom–field system may be written as:

$$|\Psi(0)_{AF}\rangle = |\Psi(0)_A\rangle \otimes |\Psi(0)_F\rangle, \tag{23}$$

We suppose that the atom is initially in the state $|\Psi(0)\rangle_{atom} = A|1\rangle + B|2\rangle + C|3\rangle$ and the initial state of light is $\sum_{n_L, n_R} g_{n_L} g_{n_R} |n_L, n_R\rangle$. Thus state vector at time $t$ is of the following form:

$$|\Psi(t)_{AF}\rangle = \hat{U}_I(t)\hat{U}_{II}(t)|\Psi(0)_{AF}\rangle, \tag{24}$$

$$|\Psi(t)_{AF}\rangle = \sum_{n_L, n_R} e^{-i(\omega_L N_1 + \omega_R N_2 + \omega_2 - (\omega_L + \omega_R))t} (A(n_L, n_R, t)|1, n_L+1, n_R\rangle + B(n_L, n_R, t)|2, n_L, n_R\rangle + C(n_L, n_R, t)|3, n_L, n_R+1\rangle) \tag{25}$$

where,

$$A(n_L, n_R, t) = A(|a_1|^2 e^{-iE_1 t} + |a_2|^2 e^{-iE_2 t} + |a_3|^2 e^{-iE_3 t})g_{n_L+1}g_{n_R}$$
$$+ B(a_1 b_1^* e^{-iE_1 t} + a_2 b_2^* e^{-iE_2 t} + a_3 b_3^* e^{-iE_3 t})g_{n_L}g_{n_R}$$
$$+ C(a_1 c_1^* e^{-iE_1 t} + a_2 c_2^* e^{-iE_2 t} + a_3 c_3^* e^{-iE_3 t})g_{n_L}g_{n_R+1}$$

$$B(n_L, n_R, t) = A(a_1^* b_1 e^{-iE_1 t} + a_2^* b_2 e^{-iE_2 t} + a_3^* b_3 e^{-iE_3 t})g_{n_L+1}g_{n_R}$$
$$+ B(|b_1|^2 e^{-iE_1 t} + |b_2|^2 e^{-iE_2 t} + |b_3|^2 e^{-iE_3 t})g_{n_L}g_{n_R}$$
$$+ C(b_1 c_1^* e^{-iE_1 t} + b_2 c_2^* e^{-iE_2 t} + b_3 c_3^* e^{-iE_3 t})g_{n_L}g_{n_R+1}$$

$$C(n_L, n_R, t) = A(a_1^* c_1 e^{-iE_1 t} + a_2^* c_2 e^{-iE_2 t} + a_3^* c_3 e^{-iE_3 t})g_{n_L+1}g_{n_R}$$
$$+ B(b_1^* c_1 e^{-iE_1 t} + b_2^* c_2 e^{-iE_2 t} + b_3^* c_3 e^{-iE_3 t})g_{n_L}g_{n_R} \tag{26}$$
$$+ C(|c_1|^2 e^{-iE_1 t} + |c_2|^2 e^{-iE_2 t} + |c_3|^2 e^{-iE_3 t})g_{n_L}g_{n_R+1}$$

## 3. The Evolution of Entropy and Atom-photon Entanglement

Defining useful measures of the entanglement for general multipartite systems is a complicated problem and an interesting topic of research at the present time [48]. For the atom–field systems, there is a unique measure of the entanglement under very general conditions such as Schmidt decomposition, local invariance, continuity and additivity). This unique measure is the quantum or von Neumann entropy of reduced density operators. The von Neumann entropy S of a system in the quantum state $\rho$ is defined as

$$S = -Tr\rho \ln \rho, \qquad (27)$$

This is the embodiment of total statistical properties of the quantum system as well as a powerful tool for the investigation of the time evolution and dynamical properties of the system. It obviously vanishes for any pure state and it is non-zero, $S \neq 0$, for mixed states. In a bipartite quantum system, a system such as atom $A$ and field $F$, and subsystem entropies at the time $t$, satisfy an important inequality due to Araki and Lieb [49]:

$$|S_A(t) - S_F(t)| \leq S_{AF}(t) \leq |S_A(t) + S_F(t)|, \qquad (28)$$

where $S_{AF}$ is the total entropy of the bipartite system and

$$S_{A(F)}(t) = -Tr[\rho_{A(F)}(t) \ln \rho_{A(F)}(t)] \qquad (29)$$

are partial entropies corresponding to reduced density operators. Based on Equation (28), for a closed atom–field system initially in a disentangled pure state, the partial entropies of the field and the atom will be equal at all times after the interaction of the two subsystems are switched on. Then our information about any of the subsystems is an indication of the entanglement of the whole system. A decreasing partial entropy means that each subsystem evolves towards a pure quantum state, whereas in an initially pure system an increasing partial entropy drives the two components to lose their individuality and become entangled [50]. So the degree of entanglement (DEM) for the atom–field system would be

$$DEM(t) = S_A = S_F = -\left(\sum_{j=1}^{3} \varepsilon_j \ln \varepsilon_j\right) \qquad (30)$$

where $\varepsilon_j$ are the eigenvalues of reduced density matrix of the atom. Now recall the system we studied in section 2. The state vector $|\Psi(t)_{AF}\rangle$ of our atom–field system at the time $t$ was given for different schemes by equations (25). Taking the partial trace of $\rho_{AF}$ over all field variables, the reduced density matrix of the atom is obtained as,

$$\rho_A = Tr_F(\rho_{AF}) = \begin{pmatrix} \rho_{11} & \rho_{12} & \rho_{13} \\ \rho_{21} & \rho_{22} & \rho_{23} \\ \rho_{31} & \rho_{32} & \rho_{33} \end{pmatrix} \qquad (31)$$

Matrix elements of the atomic reduced operator at any time can be derived as

$$\rho_{11} = \sum_{n_L,n_R} |A(n_L,n_R,t)|^2,$$

$$\rho_{22} = \sum_{n_L,n_R} |B(n_L,n_R,t)|^2,$$

$$\rho_{33} = \sum_{n_L,n_R} |C(n_L,n_R,t)|^2,$$

$$\rho_{12} = \sum_{n_1,n_2} A(n_L,n_R,t)B^*(n_L+1,n_R,t)e^{i\omega_L t}, \quad (32)$$

$$\rho_{13} = \sum_{n_L,n_R} A(n_L,n_R,t)C^*(n_L+1,n_R-1,t)e^{-i(\omega_L-\omega_R)t}$$

$$\rho_{23} = \sum_{n_L,n_R} B(n_L,n_R+1,t)C^*(n_L,n_R-1,t)e^{-i\omega_R t},$$

The eigenvalues $\varepsilon_j$ of the atomic density matrix is obtained by diagonalizing the equation (31). Based on the analytical solution presented in the previous section, we shall show the effects of different parameters on the DEM. We assume that the atom is initially in one of its states and the field is in either (I) a coherent state, or (II) a binomial state. The coherent state $|\alpha\rangle = \sum_{n=0}^{\infty} g_n |n\rangle$ is the most classical quantum state of the light with a Poissonian photon distribution. Note that $g_n$, the probability amplitude of $n$th Fock state, is given by:

$$g_n = (\alpha^n/\sqrt{n!})e^{-|\alpha|^2/2} \quad (33)$$

where $|\alpha|^2$ is the mean photon number of the coherent field. The other initial cavity field is a binomial state $|\eta,M\rangle = \sum_{n=0}^{M} g_n |n\rangle$ with

$$g_n = \left[\binom{M}{n}\eta^n(1-\eta)^{M-n}\right]^{1/2}, \quad (34)$$

In equation (34), $\eta$ and $(1-\eta)$ are the probabilities of the two possible outcomes of a Bernoulli trial. Therefore $|g_n|^2$ is the probability for having of $n$ photons when there are $M \geq n$ independent ways to produce a photon. The mean photon number in a binomial state is $\eta M$.

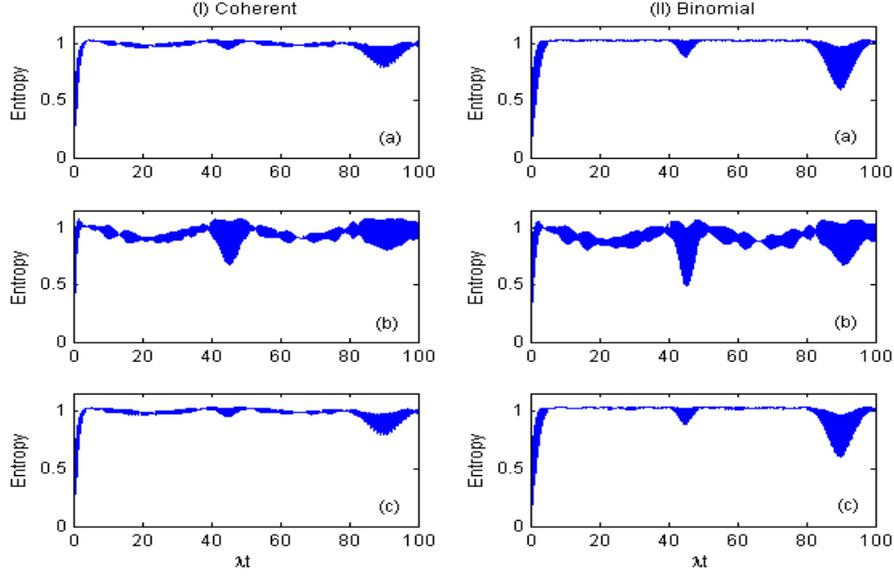

Figure 2

In the following discussion it is assumed that $\chi_R = \chi_L = 0$, that no Kerr effect is included and $\lambda_L = \lambda_R = \lambda = 1$. We consider the atom-photon DEM, when the quantized fields are in resonance with their transitions ($\Delta_L = \Delta_R = 0$). Figure 2 illustrates the evolution of the atomic entropy in terms of the scaled time $\lambda t$ in the absence of nonlinearities for a coherent (column I) and a binomial field (column II) in an intensity-independent model. We assume that the electron is initially in one of the levels. The values of the parameters are taken as follows: $|\alpha|^2 = 25$ (column I), $\eta = 0.5$  $M = 50$ (column II), $\chi_c = 0$, $\Omega = 0$ and $\beta = 0$. The figure corresponds to the initial states (a) $|1\rangle$, (b) $|2\rangle$ and (c) $|3\rangle$. We find that when the electron is in the excited state (Figure 2-b), the fluctuations in DEM are more pronounced. Moreover, for equal values of intensities, the amplitude of the fluctuations in coherent state is slightly less than that of binomial state.

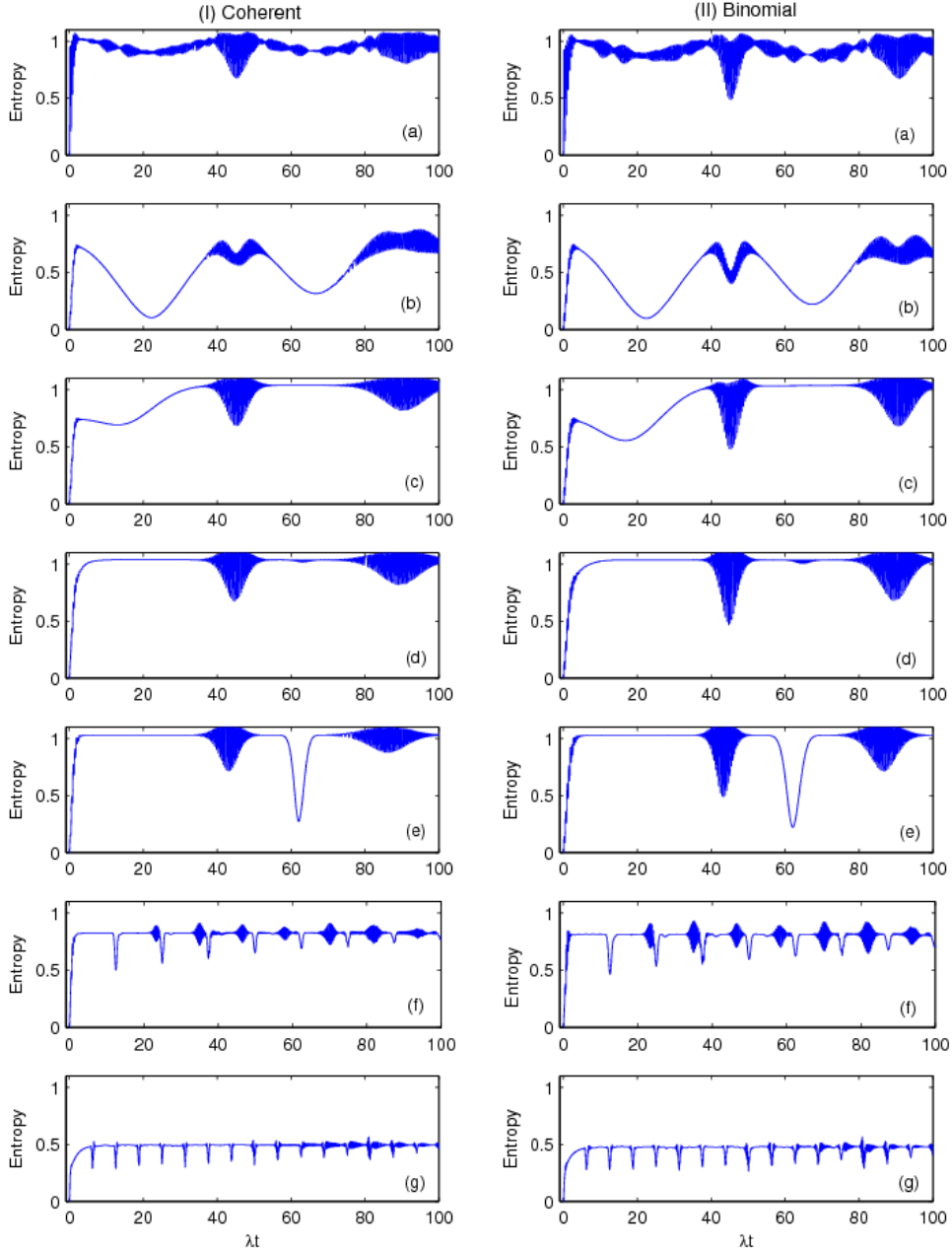

Figure 3

From now on, we assume that the atom is initially in state $|2\rangle$. Figure 3 shows the dynamical behavior of the atom-photon DEM in the presence of the cross-Kerr effect in the standard model ($\beta = 0$) for the coherent (column I) and binomial (column II) state. We note, in Figure 3, that the evolution of the atomic entropies for the cases (I) and (II) are very similar in the whole range of $0 \leq \chi_c \leq 1$ although the fluctuation amplitudes for the coherent state is slightly less than that of the binomial state. Furthermore, by including the cross-Kerr nonlinearity the collapse phenomena appears in dynamic evolution of the DEM. Note that for $\chi_c = 1$ (Figure 3-g) a considerable reduction is obtained in the maximum atomic entropy.

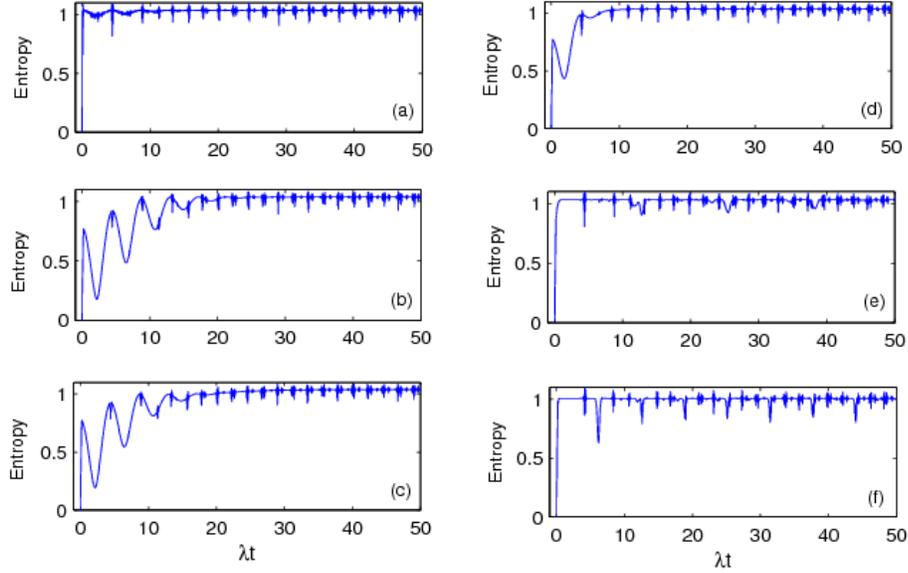

Figure 4

In the rest of the paper, it is assumed that the two quantized modes are initially in the coherent state. The situation is different for the two intensity-dependent models $\beta = 1/2$ and $\beta = -1/2$. Figure 4 displays the effect of the cross-Kerr term on the dynamical behavior of the atom-photon DEM for intensity-dependent model with $\beta = 1/2$. Here only a coherent state with $|\alpha|^2 = 25$ is assumed for the initial state of the quantized fields. It is obvious that the cross-Kerr effect, does not affect the long time DEM behavior.

In figure 5 we depict time evolution of the atom-photon DEM for the intensity dependent model with $\beta = -1/2$ when the quantized fields are initially in the coherent states. Here we see an unusual behavior as cross-Kerr coefficient is increased. It is clear that in the absence of the cross-Kerr effect, the DEM changes periodically with a tiny amplitude. By including the cross-Kerr effect, the periodic behavior of the DEM is disturbed and at the same time the mean value of the atom-photon entanglement is increased. For the larger values of $\chi_c$, however, the DEM decreases and disappears for $\chi_c = 1$.

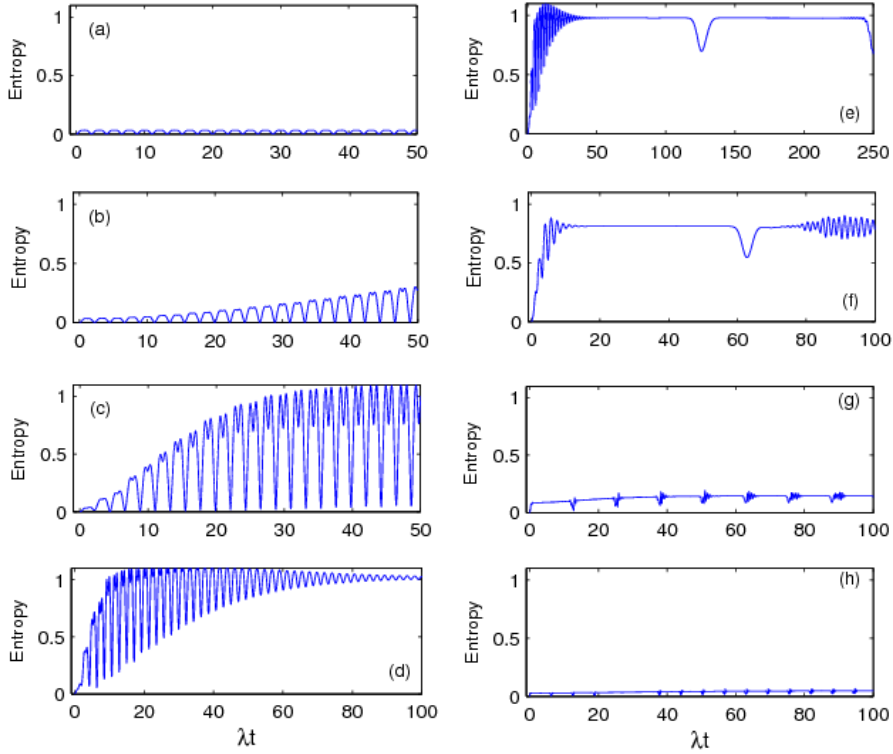

Figure 5

An interesting result is the case $\chi_c = 0.05$, in which the entropy dramatically grows to the nonzero wide flat steady state region. Comparing Figures 3, 4 and 5, one notices that the atom and the photon remain entangled for $\beta = 1/2$ and $\beta = 0$ in the whole range of $0 \leq \chi_c \leq 1$, while here for the case of $\beta = -1/2$, it drops to negligible values for larger values of $\chi_c$.

Now, we turn our attention to the classical driving field and investigate its effect on the time evolution of the atom-photon DEM for (I) $\beta = 0$ and (II) $\beta = 1/2$. Figure 6 shows our numerical DEM results for the following parameters: $|\alpha|^2 = 20$, $\chi_c = 0$. Note that $\Omega$ is compared by $\sqrt{|\alpha|^2 + 1}$ and $|\alpha|^2 + 1$ in the model with $\beta = 0$ and $\beta = 1/2$, respectively. An investigation on figure 6 shows that by increasing the intensity of the classical driving field, the DEM in the both models decreases and the system eventually becomes disentangled for $\beta = 0$. The physical interpretation lies in the population distribution in quantum levels. Our numerical calculations show that while the intensity of the classical driving field is increased, the upper level remains more populated, which leads to a negligible residual atom-photon entanglement for high values of $\Omega$. Similar behavior is observed for $\beta = -1/2$ and the disentanglement occurs at the lower intensities compared to $\beta = 0$.

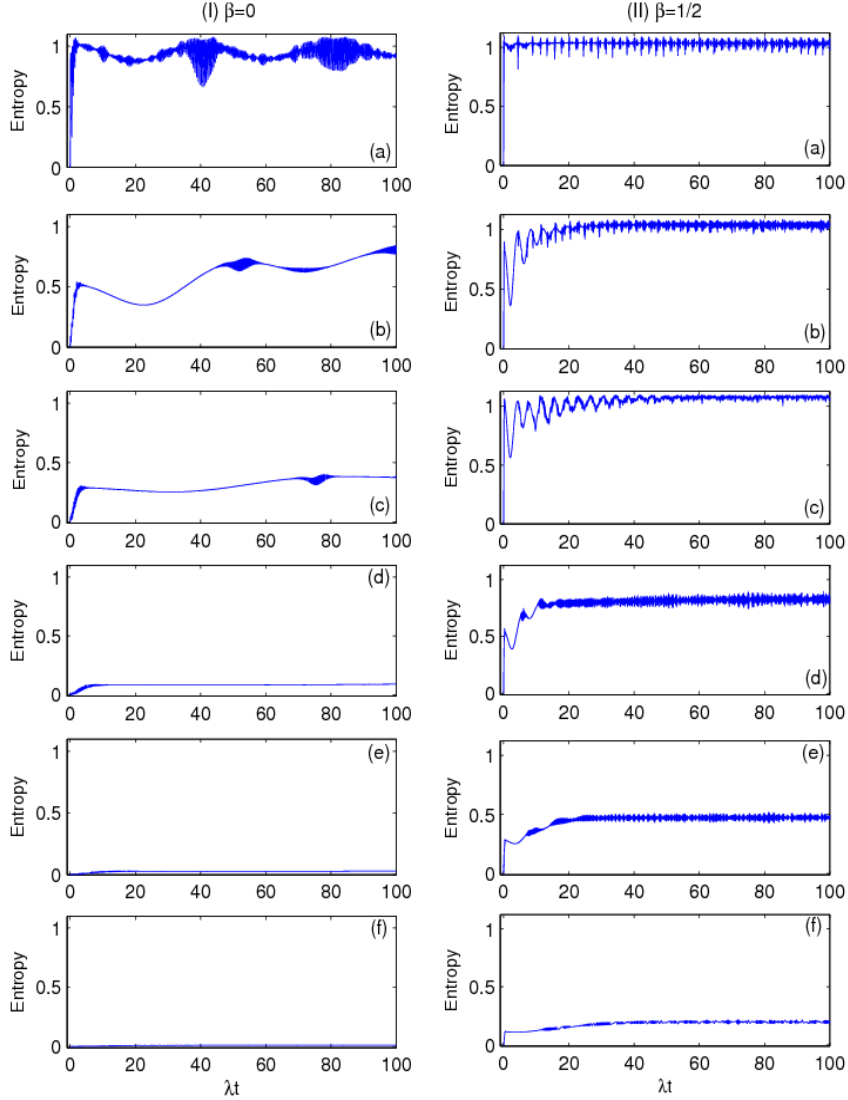

Figure 6

**4. Two-mode Entanglement**

In this section we will describe the dynamical behavior of the quantum entanglement between two quantized modes of the cavity radiation following their interaction with the atom. The two modes are assumed to be initially in the coherent states. To this goal, we need an entanglement criterion for continuous variables corresponding to the states of infinite-dimensional systems. We apply the criterion introduced by Duan, Giedke, Cirac and Zoller (DGCZ) [28] which, in general, is a sufficient condition to establish bipartite continuous-variable entanglement. It is well-known that the position and momentum (continuous variables) of the quantized electromagnetic modes are represented by quadrature operators of quantum harmonic oscillators defined as

$$\hat{x}_j = (\hat{a}_j^+ + \hat{a}_j)/\sqrt{2}, \quad \hat{p}_j = i(\hat{a}_j^+ - \hat{a}_j)/\sqrt{2} \quad (j = R, L). \tag{35}$$

Then a pair of EPR-type operators is defined as $\hat{u} = \hat{x}_R + \hat{x}_L$, $\hat{v} = \hat{p}_R - \hat{p}_L$ [28]. According to DGCZ, two quantized modes of our cavity radiation are entangled if the quantum fluctuations of the operators $\hat{v}$ and $\hat{u}$ satisfy the inequality

$$\Delta\hat{u}^2 + \Delta\hat{v}^2 = \left\langle (\Delta\hat{u})^2 + (\Delta\hat{v})^2 \right\rangle < 2. \tag{36}$$

Substituting the operators $\hat{v}$ and $\hat{u}$ into Equation (36), the total variance of the operators $\hat{u}$ and $\hat{v}$ is obtained in term of the operators $\hat{a}_j$, $\hat{a}_j^+$ as

$$\Delta\hat{u}^2 + \Delta\hat{v}^2 = \left\langle (\Delta\hat{u})^2 + (\Delta\hat{v})^2 \right\rangle = 2[1 + <\hat{a}_R^+\hat{a}_R> + <\hat{a}_L^+\hat{a}_L> + <\hat{a}_R^+\hat{a}_L^+> + <\hat{a}_R\hat{a}_L> - \\ <\hat{a}_R><\hat{a}_R^+> - <\hat{a}_L^+><\hat{a}_L> - <\hat{a}_R^+><\hat{a}_L^+> - <\hat{a}_R><\hat{a}_L>] \tag{37}$$

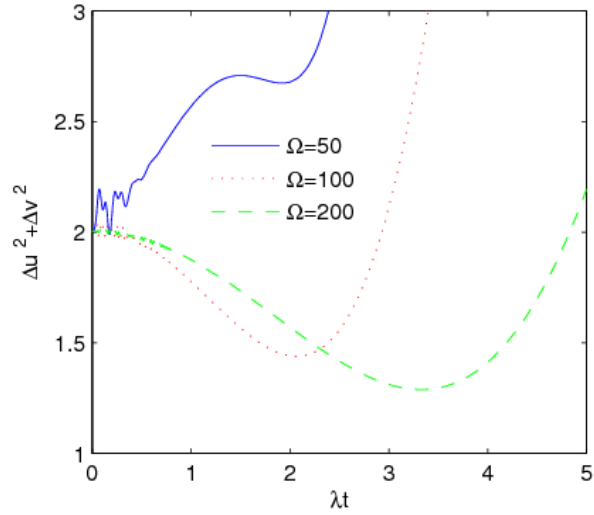

Figure 7

Our calculations show that the two-mode entanglement occurs only for the intensity-dependent coupling model with $\beta = 1/2$. In Figure 7, the total variance $\Delta\hat{u}^2 + \Delta\hat{v}^2$ is plotted in terms of the scaled time $\lambda t$ in the absence of cross-Kerr nonlinearity for different Rabi frequency of classical driving fields. The other parameters are chosen as $|\alpha|^2 = 20$, $\Delta_R = \Delta_L = 0$. In such circumstances, the two-mode entanglement emerges for considerable amount of classical field intensity ($\Omega > 50$). It can be seen that in short times, as the intensity of the classical field is increased, the duration of the two-mode entanglement grows longer. There is no entanglement of the two-mode in long times.

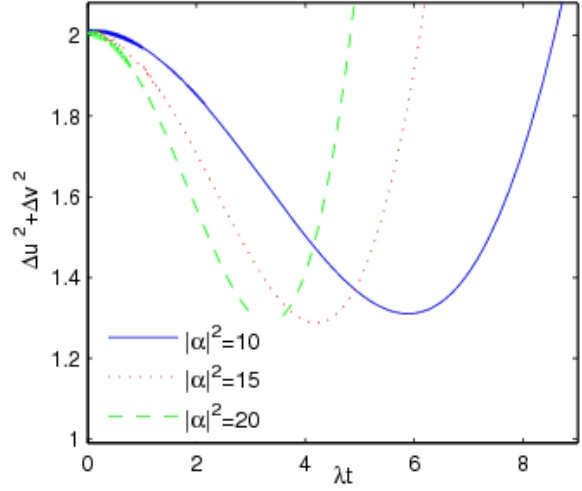

Figure 8

Figure 8 shows the effect of field intensity on the time evolution of total variance $\Delta\hat{u}^2 + \Delta\hat{v}^2$. The other parameters are the same as in the dashed green line curve of figure 7. It is clear that an increase in the field intensity results in a decrease in the duration of the entanglement. The modes disentangle earlier in time.

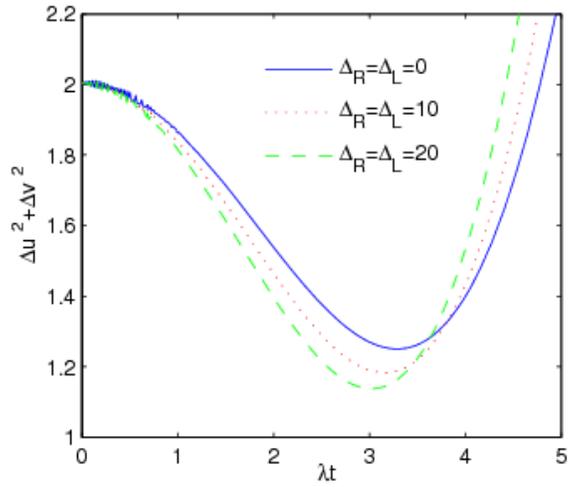

Figure 9

The time evolution of total variance $\Delta\hat{u}^2 + \Delta\hat{v}^2$ for different values of detuning parameters is shown in figure 9. The other considered parameters are as the green dashed line in figure 8. It is easy to see that the longest time that the two-mode entanglement will be achieved corresponds to the one-photon resonance condition ($\Delta_L = \Delta_L = 0$). Beyond one-photon resonance condition, the time duration of the two-mode entanglement decreases.

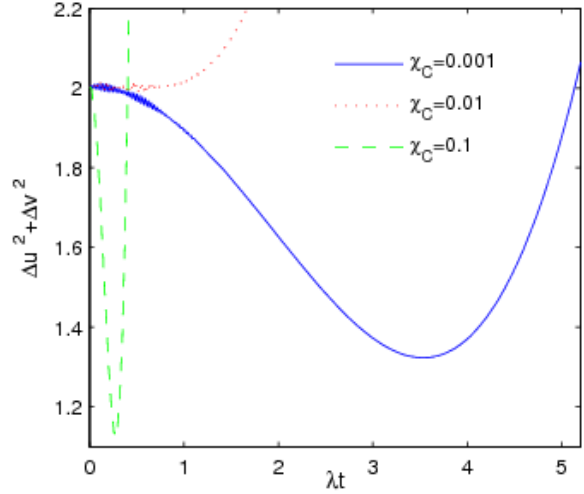

Figure 10

Figure 10 displays the effect of the cross-Kerr coefficient on the dynamical behavior of the total variance $\Delta\hat{u}^2 + \Delta\hat{v}^2$ for the one-photon resonance condition. The other parameters are the same as the solid blue curve in figure 9. We show that by increasing the cross-Kerr coefficient, the two-mode entanglement is destroyed, though, it is again appeared just for a short time in the beginning of interaction.

**5. Two-mode quadrature squeezing**

The light produced by any optical system is an excitation of various modes of electromagnetic fields. In the two-photon devices, the output light is generated by the simultaneous emissions of two photons into two of the output modes [51, 52]. One of the most important two-photon devices is the optical parametric oscillators (OPOs) which are well-known sources of the nonclassical light [53]. Degenerate and non-degenerate OPOs have been used to generate the single mode [54] and the two-mode squeezed states [55]. The output of a non-degenerate OPO is an example of a two-mode squeezed state. There are two orthogonal quadrature components corresponding to the two modes. The squeezing takes place not in the individual modes but in the total quadrature components. In quantum optics, the squeezed states are recognized by the property that the quantum fluctuations in one of the field quadratures are smaller than those of the coherent light or the vacuum. The definition of the single-mode squeezing can be generalized to investigate squeezing produced by the two-mode interaction [51,52,56]. In 1980s and early 1990s much works was done to study the two-mode squeezing of the cavity fields interacting with two-level and three-level atoms [57-59].

Here we study the dynamical behavior of quadrature squeezing of the two-mode quantized radiation in our model. These properties of the two-mode cavity quantized radiation can be described by two quadrature operators defined as

$$\hat{c}_+ = \hat{x}_R + \hat{x}_L, \qquad \hat{c}_- = \hat{p}_R + \hat{p}_L, \qquad (38)$$

which satisfy the commutation relation $[\hat{c}_+, \hat{c}_-] = 2i$. A two-mode field is said to be in a two-mode squeezed state if either $\Delta \hat{c}_+^2 < 1$ or $\Delta \hat{c}_-^2 < 1$. The variances of the quadrature operators can be expressed as

$$\Delta \hat{c}_\pm^2 = <\hat{c}_\pm^2> - <\hat{c}_\pm>^2. \qquad (39)$$

Substituting equation (38) in equation (39), one can obtain the variances as

$$\Delta \hat{c}_\pm^2 = 1 + <\hat{a}_R^+ \hat{a}_R> + <\hat{a}_L^+ \hat{a}_L> + <\hat{a}_R^+ \hat{a}_L> + <\hat{a}_L^+ \hat{a}_R> \pm \frac{1}{2}(<\hat{a}_R^{+2}> + <\hat{a}_L^{+2}> + <\hat{a}_R^2> + <\hat{a}_L^2> + 2<\hat{a}_R \hat{a}_L> + 2<\hat{a}_R^+ \hat{a}_L^+>). \qquad (40)$$

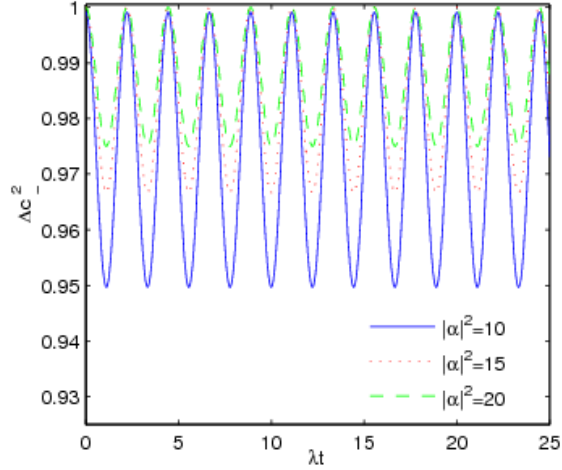

Figure 11

Our calculations show that only for $\beta = -1/2$ and in the absence of a classical driving field, the two-mode squeezing occurs. The time evolution of the two-mode squeezing criterion for different intensities of the two-mode quantized modes is shown in figure 11. The other parameters are $\chi_c = 0$, $\Delta_R = \Delta_L = 0$. Figure 11 shows that the two-mode squeezing in our system has oscillatory dynamical behavior in which the frequency of the oscillations does not depend on the intensity of the two-mode quantized field. However, by increasing the field intensity, the maximum two-mode squeezing is reduced.

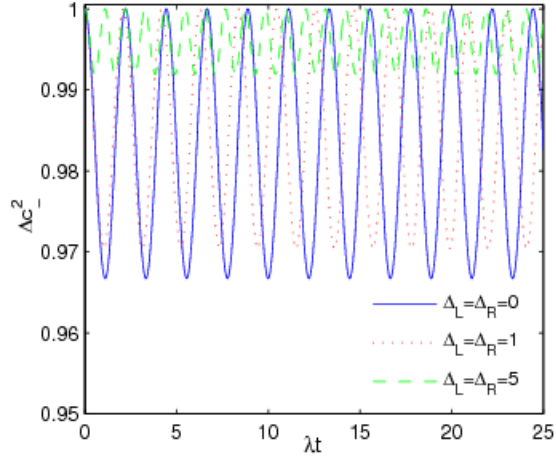

Figure 12

Detuning of the two-mode quantized field is another parameter for controlling the two-mode squeezing. In figure 12, we plot the dynamical behavior of $\Delta\hat{c}_-^2$ for different values of detuning parameters. The other parameters are the same as the dotted red line in figure 11. It is seen that the oscillation frequency depends on the detuning of the two-mode quantized field. Moreover, the period and the maximum two-mode squeezing is decreased when the detuning of the quantized field is increased.

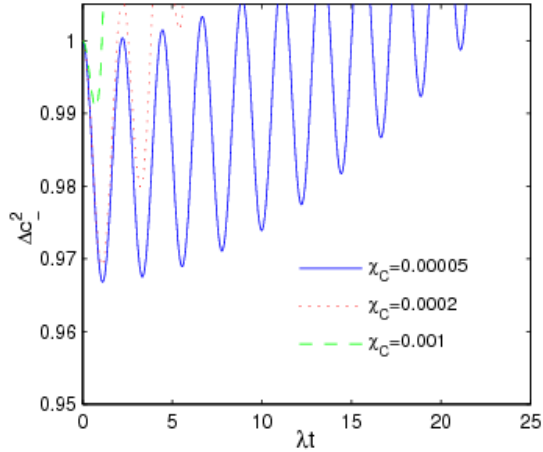

Figure 13

We, finally, include the cross-Kerr effect on the dynamical behavior of $\Delta\hat{c}_-^2$. Figure 13 displays $\Delta\hat{c}_-^2$ in terms of the scaled time for different values of the cross-Kerr coefficient. The other used parameters are the same as the solid blue line in Figure 12. It is evident that the long-time two-mode squeezing disappears as the cross-Kerr coefficient increases beyond a certain normal value limit. One can, therefore, deduce that the two-mode squeezing in this system is very sensitive to cross-Kerr effect.

## 6. Conclusions

The effect of cross Kerr nonlinearity on the entanglement of atom and fields was investigated in a $\Lambda$-type atomic system. It was shown that the dynamical behavior of the entanglement depends on the cross-Kerr coefficient of the medium. Moreover, the two-mode entanglement was established by applying a classical field to the lower levels. It was deduced that the two-mode squeezing can be generated in the absence of cross-Kerr nonlinearity. Our results showed that the cross-Kerr nonlinearity has a destructive role in the two-mode entanglement (squeezing).

**Acknowledgment**

Ali Mortezapour would like to Dr. Bahman Ahmadi for his useful assistance in English writing of this paper.

**Figures caption**

**Figure 1.** A $\Lambda$-type three level atom coupled with a two-mode field of frequencies $\omega_L$ and $\omega_R$ with detuning $\Delta_L$, $\Delta_R$ and a classical driving field that couples the lower two levels with the Rabi frequency $\Omega$.

**Figure 2.** Dynamical behavior of the atomic reduced entropy when the field is (I) Coherent($|\alpha|^2 = 25$) and (II) Binomial($\eta = 0.5$, $M = 50$) with the following parameters: $\Omega = 0$, $\chi_c = 0$, and $\beta = 0$ for (a) $A = 1, B = C = 0$, (b) $A = C = 0, B = 1$ and (c) $A = B = 0$, $C = 1$.

**Figure 3.** Dynamical behavior of the atomic reduced entropy when field is (I) Coherent($|\alpha|^2 = 25$) and (II) Binomial($\eta = 0.5$, $M = 50$) with the following parameters: $A = C = 0, B = 1$, $\Omega = 0$, and $\beta = 0$ for (a) $\chi_C = 0$, (b) $\chi_C = 0.001$, (c) $\chi_C = 0.01$, (d) $\chi_C = 0.05$, (e) $\chi_C = 0.1$, (f) $\chi_C = 0.5$, (g) $\chi_C = 1$.

**Figure 4.** Dynamical behavior of the atomic reduced entropy when the initial field state is Coherent with $|\alpha|^2 = 25$ for the intensity dependent model with $\beta = 1/2$ when the other parameters are: $\Omega = 0$ and for (a) $\chi_C = 0$, (b) $\chi_C = 0.001$, (c) $\chi_C = 0.01$, (d) $\chi_C = 0.05$ and (e) $\chi_C = 0.1$, (f) $\chi_C = 1$.

**Figure 5.** Dynamical behavior of the atomic reduced entropy when the initial field state is Coherent with $|\alpha|^2 = 25$ for the intensity dependent model with $\beta = -1/2$ when the other parameters are: $\Omega = 0$ and for (a) $\chi_C = 0$, (b) $\chi_C = 0.002$, (c) $\chi_C = 0.01$, (d) $\chi_C = 0.03$, (e) $\chi_C = 0.05$, (f) $\chi_C = 0.1$, (g) $\chi_C = 0.5$, (h) $\chi_C = 1$.

**Figure 6.** Dynamical behavior of the atomic reduced entropy when the initial field state is Coherent with $|\alpha|^2 = 20$ for $\beta = 0$. Other used parameters are: $\chi_C = 0$, for (a) $\Omega = 0$, (b) $\Omega = 10$, (c) $\Omega = 20$, (d) $\Omega = 50$, (e) $\Omega = 100$, (f) $\Omega = 200$.

**Figure 7.** Plots of the total variance $\Delta\hat{u}^2 + \Delta\hat{v}^2$ in terms of $\lambda t$ for different Rabi frequency of classical driving field in the intensity dependent model with $\beta = 1/2$ and when the other

parameters are: $\chi_c = 0$, $|\alpha|^2 = 20$, $\Delta_L = \Delta_R = 0$, and $\Omega = 50$ (Solid blue line), $100$ (Dotted red line), $200$ (Dashed green line)

**Figure 8.** Plots of total variance $\Delta\hat{u}^2 + \Delta\hat{v}^2$ in terms of $\lambda t$ for the different intensity of cavity radiations in the intensity dependent model with $\beta = 1/2$ and when the other parameter are: $\Omega = 200$, $\chi_c = 0$, $\Delta_L = \Delta_R = 0$, $|\alpha|^2 = 10$ (Solid blue line), $15$ (Dotted red line), $20$ (Dashed green line).

**Figure 9.** Plots of the total variance $\Delta\hat{u}^2 + \Delta\hat{v}^2$ in terms of $\lambda t$ for different values of detuning parameters in the intensity dependent model with $\beta = 1/2$ and when the other parameter are: $|\alpha|^2 = 20$, $\Omega = 200$, $\chi_c = 0$, $\Delta_L = \Delta_R = 0$ (Solid blue line), $10$ (Dotted red line), $20$ (Dashed green line).

**Figure 10.** Plots of total variance $\Delta\hat{u}^2 + \Delta\hat{v}^2$ in terms of $\lambda t$ for different values of cross-Kerr coefficient in the intensity dependent model with $\beta = 1/2$ and when the other parameter are: $|\alpha|^2 = 20$, $\Omega = 200$, $\Delta_L = \Delta_R = 0$, $\chi_c = 0.001$ (Solid blue line), $0.01$ (Dotted red line), $0.1$ (Dashed green line).

**Figure 11.** Dynamical behavior of the two-mode squeezing measure of cavity fields for the different intensity of cavity fields in the intensity dependent model with $\beta = -1/2$ and when the other parameter are: $\Omega = 0$, $\chi_c = 0$, $\Delta_L = \Delta_R = 0$, $|\alpha|^2 = 10$ (Solid blue line), $15$ (Dotted red line), $20$ (Dashed green line).

**Figure 12.** Dynamical behavior of the two-mode squeezing measure of cavity fields for the different values of detuning parameters in the intensity dependent model with $\beta = -1/2$ and when the other parameter are: $|\alpha|^2 = 15$, $\Omega = 0$, $\chi_c = 0$, $\Delta_L = \Delta_R = 0$ (Solid blue line), $1$ (Dotted red line), $5$ (Dashed green line).

**Figure 13.** Dynamical behavior of the two-mode squeezing measure of cavity fields for the different values of cross-Kerr coefficient in the intensity dependent model with $\beta = -1/2$ and when the other parameter are: $|\alpha|^2 = 15$, $\Delta_L = \Delta_R = 0$, $\chi_c = 0.00005$ (Solid blue line), $0.0002$ (Dotted red line), $0.001$ (Dashed green line).